\documentstyle[ific_title,wstwocl1a,epsfig]{article}
\begin{document}
\bibliographystyle{unsrt}    
\newcommand{\st}{\scriptstyle}
\newcommand{\sst}{\scriptscriptstyle}
\newcommand{\mco}{\multicolumn}
\newcommand{\epp}{\epsilon^{\prime}}
\newcommand{\vep}{\varepsilon}
\newcommand{\ra}{\rightarrow}
\newcommand{\ppg}{\pi^+\pi^-\gamma}
\newcommand{\vp}{{\bf p}}
\newcommand{\ko}{K^0}
\newcommand{\kb}{\bar{K^0}}
\newcommand{\al}{\alpha}
\newcommand{\ab}{\bar{\alpha}}
\def\be{\begin{equation}}
\def\ee{\end{equation}}
\def\bea{\begin{eqnarray}}
\def\eea{\end{eqnarray}}
\def\CPbar{\hbox{{\rm CP}\hskip-1.80em{/}}}

\def\ap#1#2#3   {{\em Ann. Phys. (NY)} {\bf#1} (#2) #3}
\def\apj#1#2#3  {{\em Astrophys. J.} {\bf#1} (#2) #3}
\def\apjl#1#2#3 {{\em Astrophys. J. Lett.} {\bf#1} (#2) #3}
\def\app#1#2#3  {{\em Acta. Phys. Pol.} {\bf#1} (#2) #3}
\def\ar#1#2#3   {{\em Ann. Rev. Nucl. Part. Sci.} {\bf#1} (#2) #3}
\def\cpc#1#2#3  {{\em Computer Phys. Comm.} {\bf#1} (#2) #3}
\def\err#1#2#3  {{\it Erratum} {\bf#1} (#2) #3}
\def\ib#1#2#3   {{\it ibid.} {\bf#1} (#2) #3}
\def\jmp#1#2#3  {{\em J. Math. Phys.} {\bf#1} (#2) #3}
\def\ijmp#1#2#3 {{\em Int. J. Mod. Phys.} {\bf#1} (#2) #3}
\def\jetp#1#2#3 {{\em JETP Lett.} {\bf#1} (#2) #3}
\def\jpg#1#2#3  {{\em J. Phys. G.} {\bf#1} (#2) #3}
\def\mpl#1#2#3  {{\em Mod. Phys. Lett.} {\bf#1} (#2) #3}
\def\nat#1#2#3  {{\em Nature (London)} {\bf#1} (#2) #3}
\def\nc#1#2#3   {{\em Nuovo Cim.} {\bf#1} (#2) #3}
\def\nim#1#2#3  {{\em Nucl. Instr. Meth.} {\bf#1} (#2) #3}
\def\np#1#2#3   {{\em Nucl. Phys.} {\bf#1} (#2) #3}
\def\pcps#1#2#3 {{\em Proc. Cam. Phil. Soc.} {\bf#1} (#2) #3}
\def\pl#1#2#3   {{\em Phys. Lett.} {\bf#1} (#2) #3}
\def\prep#1#2#3 {{\em Phys. Rep.} {\bf#1} (#2) #3}
\def\prev#1#2#3 {{\em Phys. Rev.} {\bf#1} (#2) #3}
\def\prl#1#2#3  {{\em Phys. Rev. Lett.} {\bf#1} (#2) #3}
\def\prs#1#2#3  {{\em Proc. Roy. Soc.} {\bf#1} (#2) #3}
\def\ptp#1#2#3  {{\em Prog. Th. Phys.} {\bf#1} (#2) #3}
\def\ps#1#2#3   {{\em Physica Scripta} {\bf#1} (#2) #3}
\def\rmp#1#2#3  {{\em Rev. Mod. Phys.} {\bf#1} (#2) #3}
\def\rpp#1#2#3  {{\em Rep. Prog. Phys.} {\bf#1} (#2) #3}
\def\sjnp#1#2#3 {{\em Sov. J. Nucl. Phys.} {\bf#1} (#2) #3}
\def\spj#1#2#3  {{\em Sov. Phys. JEPT} {\bf#1} (#2) #3}
\def\spu#1#2#3  {{\em Sov. Phys.-Usp.} {\bf#1} (#2) #3}
\def\zp#1#2#3   {{\em Zeit. Phys.} {\bf#1} (#2) #3}

\setcounter{secnumdepth}{2} 
\begin{ificpage}
     \IFICname
     \docnum{IFIC/95--60}
     \hepdocnum{hep--ex/9511002}
     \date{}
     \vspace*{0.5cm}
\titol{Charged Particle Production in the \\
       Fragmentation of Quark and Gluon Jets}
     \vspace*{0.5cm}
\author{\begin{center}J. Fuster~\instref{ific} and S. Mart\'{\i}~\instref{ific}
\end{center} }
     \instfoot{ific}{~~~IFIC, centre mixte CSIC--Universitat de Val\`encia,
               Avda. Dr. Moliner 50, E--46100 Burjassot, Val\`encia, Spain}  
     \vspace*{2.0cm}

     \begin{shortabstract}
{\large
Recent results on the total production and angular distribution of charged
particles o\-ri\-gi\-na\-ted from the fragmentation of quark and gluon jets are
presented. Experimental studies of the multiplicity as a function of the quark
and gluon jet energy, the inter-jet particle flow and the individual
fragmentation functions are reviewed and compared to expectations from QCD.}

     \end{shortabstract}
     \vspace*{5cm}
     \docinfo{To appear in the proceedings of the XXXII conference of the
     European Physics Society, \\ 
     High Energy Physics, Brussels, 27 July -- 3 August, 1995.}
 \end{ificpage}

   
\title{CHARGED PARTICLE PRODUCTION IN THE FRAGMENTATION OF \\
QUARK AND GLUON JETS}

\firstauthors{J. Fuster and S. Mart\'{\i} }
\firstaddress{IFIC, centre mixte CSIC--Universitat de Val\`encia,
Avda. Dr. Moliner 50,E--46100 Burjassot,Spain \\
e-mail: \verb+fuster@evalvx.ific.uv.es+ and \verb+martis@evalo1.ific.uv.es+}

\twocolumn[\maketitle\abstracts{Recent  results on  the total production  and
angular  distribution of charged particles  originated from the fragmentation
of   quark and   gluon  jets  are presented.    Experimental   studies of the
multiplicity as a function of the  quark and gluon  jet energy, the inter-jet
particle  flow and the individual  fragmentation  functions are reviewed  and
compared to the expectations from QCD.}]

\section{Introduction} 
In Quantum Chromodynamics  (QCD), quarks ($q$) and  gluons ($g$) are coloured
objects that carry different colour charges.  The strength of the quark-gluon
and gluon-gluon couplings is directly related with the colour charges and, in
consequence, the probability to emit  additional gluons is also different for
both quarks and  gluons.  Reconstructed quark  and gluon jets are expected to
inherit their parton origin  and, therefore, they should exhibit  differences
in their multiplicities, energies  and   angular distributions.  It  is  well
known,   however, the existing difficulties   to measure these differences in
quantitative  agreement   with the  predictions  from  perturbative   QCD, as
partons, quarks and gluons, are not directly observed  in nature and only the
stable     particles,   produced  after     the  fragmentation  process,  are
experimentally detected.  Earlier measurements of the  ratio of the gluon jet
multiplicity  w.r.t  the quark jet  multiplicity,   $r=\langle N_g \rangle  /
\langle N_q \rangle$, of the coherence phenomena  in the inter-jet $q\bar{q}$
region  w.r.t. the $qg$ region,   etc., have established qualitative evidence
for  these differences but have failed   in describing them quantitatively as
predicted by QCD.

At present, more  than 16 million hadronic decays  of the $Z$ boson have been
collected  by  the LEP experiments.    These  high statistics allow  applying
restrictive  selection  criteria to select quark   and gluon jet samples with
high purities.    The selected data  samples  are almost background  free and
small  corrections to  account for impurities   are  needed. A  smaller model
dependence than ever  is now  achieved, bringing  the possibility  to perform
quantitative   studies of  quark    and gluon    fragmentation  according  to
perturbative QCD.

Studies on quark and gluon jet fragmentation have been  carried out at LEP by
the  OPAL\cite{opal_qg} and  ALEPH\cite{aleph_qg}   collaborations  selecting
three-jet  symmetric configurations in     which the  quark  and gluon    jet
properties   were compared at   a  fixed energy  scale.   A new analysis from
DELPHI\cite{delphi_qg} using   symmetric  and non symmetric  three-jet  event
configurations show these  properties as a   function of the jet  energy. The
hadronic  part of radiative  $q \bar{q}\gamma$  events  as a  function of the
reduced  center of mass   energy, $s'=s\cdot(1-2E_\gamma/\sqrt{s})$, has also
been investigated  using   event shape variables  by L3\cite{l3_qqgamma}  and
using charged   multiplicity  distributions  by  DELPHI\cite{delphi_qg}.  All
these   studies are quite  extensive  and a   proper description covering all
details is  out of the scope  of  this short review.  Here only  results from
analyses, which  use heavy flavour  tagging   in $q\overline{q}g$  events  to
identify  gluon jets and  radiative $q\overline{q}\gamma$  events to identify
quark jets, are reported.  Other  presentations  of similar content based  on
other  type of    jet  tagging   procedures   were  also  discussed  in   the
conference\cite{ron}. The  differences of quarks and  gluons as a function of
the jet reconstruction algorithm  and of  the three-jet event  configuration:
jet energies,  the  particle  flow  in different inter-jet   regions and  the
scaling violation effects, are the subject of our discussion following.

\section{Quark and gluon jet  selection}
Gluon  and quark jets are selected  using hadronic three-jet events. Jets are
mainly  reconstructed using the  {\sc  Durham}  algorithm  (OPAL,  ALEPH, L3,
DELPHI) although the   {\sc Jade} and {\sc  Cone}  algorithms have also  been
used\cite{opal_qg,delphi_qg},  in particular to  observe  the effects due  to
different angular particle   acceptance of  the  various algorithms   and  to
compare with the results from hadron colliders.

In  the  gluon splitting process  ($g\rightarrow  q\bar{q}$), the heavy quark
production is  strongly suppressed\cite{opal_cc}.  This opens the possibility
to  select gluon jets from  $q\bar{q}g$ events in which  two of the jets, the
quark  jets,   are seen  to  satisfy the   experimental  signatures  of being
initiated by  $b$ quarks, leaving the remaining  jet to be  associated to the
gluon  jet  without further requirements.    Algorithms for tagging  $b$ jets
exploit the fact that  the decay products of long  lived B hadrons have large
impact parameters and/or contain inclusive  high momentum leptons coming from
the  semileptonic decays of the  B hadrons.  Gluon  purities of 94\% and 85\%
are achieved when using these techniques,  respectively. Obviously, the quark
jets belonging to these events cannot be used to  represent an unbiased quark
sample. Thus   the quark jets whose  properties  are to be  compared with the
gluon  jets must  be  selected from other   sources which in  any case should
preserve  the same kinematics. Two  possibilities  have been proposed in  the
current   literature.  One consists  in selecting   symmetric three-jet event
configurations\cite{opal_qg,aleph_qg,delphi_qg} in which  one (Y) or the  two
(Mercedes) quark jets have similar energy to that of the gluon jet. The quark
jet  purities reached are $\sim$48\%  and $\sim$66\%, for  Y and for Mercedes
events,         respectively.               In         a               second
solution\cite{opal_qg,delphi_qg,l3_qqgamma} radiative $q\bar{q}\gamma$ events
are selected, allowing  a  sample of quark  jets  with variable energy  to be
collected.  In this latter case,  misidentifications of $\gamma$'s due to the
$\pi^\circ$  background and radiative $\tau^+\tau^-\gamma$ contamination give
rise to quark jet  purities of $\sim$92\%.  This method gives a higher purity
but unfortunately suffers from the lack of statistics.
  
\section{Multiplicities of quark and gluon jets}

Results   on    the   charged    multiplicity     of    quark   and     gluon
jets\cite{opal_qg,aleph_qg}  using     symmetric   Y   configurations     and
reconstructed with {\sc Durham} at 24  GeV gluon jet  energy, give a ratio of
$r \approx 1.23  \pm 0.04 \mbox{(stat.+syst.)}$  which does not depend on the
cut-off parameter ($y_{cut}$) selected  to reconstruct jets\cite{opal_qg}. It
is significantly  higher than one, which  indicates that quark  and gluons in
fact fragment differently,  but it  remains far  from the  asymptotic  lowest
order expectation  of $C_F/C_A   =   9/4$,  suggesting  that higher     order
corrections and non-perturbative effects are very important to understand the
measured value.  A  next-to-leading order  correction\cite{mueller}  in  MLLA
(Modified  Leading Log Approximation) at $\cal{O}$$(\sqrt{\alpha_s})$ already
lowers the  prediction towards $r$ values  slightly below two and  exhibits a
small energy  dependence due to the  running of $\alpha_s$.  However  this is
still insufficient to explain the value of $r$ determined by the experiments.
Solutions    based   on    the   Monte    Carlo    method  give    a   better
approximation\cite{delphi_qg}.  The parton shower option  of the {\sc Jetset}
generator\cite{jetset} which uses   the Altarelli-Parisi splitting  functions
for     the   evolution  of  the   parton    shower   reduce  the theoretical
prediction\cite{delphi_qg} for $r$. At parton  level,  at 24 GeV jet  energy,
the expected value is $\sim$1.4 and it is further reduced to $\sim$1.3 if the
value of $r$ is computed after the fragmentation process. In both cases there
is a clear dependence of $r$ with the jet energy\cite{delphi_qg} which can be
parametrized using straight lines  with  slopes of  ${\Delta r}/{\Delta E}  =
(+90 \pm 3\mbox{(stat.)})\cdot 10^{-4} \ \mbox{GeV}^{-1}$ at parton level and
${\Delta   r}/{\Delta E}      =  (+76  \pm    2\mbox{(stat.)})\cdot  10^{-4}\
\mbox{GeV}^{-1}$ after fragmentation. The absolute  value of $r$ predicted at
parton level is however largely affected by the choice on the $Q_0$ parameter
(cut-off at which the parton evolution stops) but has negligible influence on
its relative  variation with the energy, i.e.  the slope. The DELPHI analysis
uses symmetric and  non-symmetric  three-jet event configurations  with quark
and gluon jets  of variable energy, allowing  thus  all these  properties and
predictions to be tested. A value of $r=1.24\pm0.03\ \mbox{(stat.+syst.)}$ is
measured corresponding to an average  jet energy of  $\sim$27 GeV. The energy
dependence of $r$ is also suggested at 2.7$\sigma$ significance level, with a
fitted slope of ${\Delta r}/{\Delta E} = (+86\pm 32\mbox{(stat.+syst.)})\cdot
10^{-4}\ \mbox{GeV}^{-1}$.

\begin{figure}[hbt]
\begin{center}
\mbox{\epsfig{file=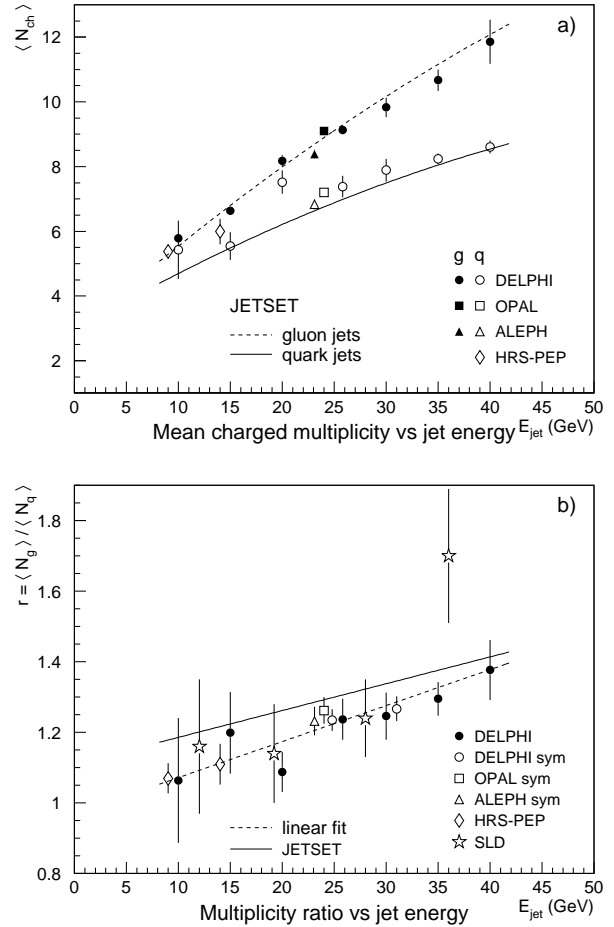,width=8.2cm}}
\end{center}
\caption{(a) Mean   charged  multiplicity of quark  and    gluon jets as  a
function   of the jet  energy. Original 
results are corrected to account for the same 
flavour composition of 11\%  $b$'s and 33\% $c$'s. 
Solid and dashed lines  represent the {\sc Jetset} event   generator 
prediction. (b) Multiplicity
ratio $r$ as a function of the jet energy. The solid line represents 
the {\sc Jetset}  prediction after fragmentation
 and  the dashed line is the best fit  to data. Values corresponding 
to the  same energy  are separated for having a better display.}
\label{fig:proc_mult}
\end{figure}

The absolute value  of $r$ depends on the  reconstruction jet algorithm.  For
both  the {\sc Jade} and {\sc   Cone} schemes different  results are obtained
w.r.t.   the {\sc Durham} scheme\cite{opal_qg,delphi_qg}. This  is due to the
combined  effect   of   the   different  sensitivity  of   the    various jet
reconstruction  algorithms to  soft  particles  at  large angles   and of the
expected different  angular and energy spectra  of the emitted soft gluons in
the quark and gluon jets.  A precise deconvolution of  both effects seems, at
present,  impossible\cite{ruso}. The energy  behaviour of $r$ is however well
represented\cite{delphi_qg}.
 
All published data from various experiments can be used  to perform a general
study in which  more precise results  can be attained. Care  has  to be taken
when  comparing  results from different  experiments as  not all the analyses
consider  quark   samples   of the   same  flavour  content,    and this  can
significantly      change     the           quark     jet        multiplicity
\cite{delphi_bm,opal_bm,sld_bm}.   To  optimize calculus,  the results  given
following are based on     the $q\bar{q}\gamma$ analysis from   DELPHI  which
considers that the quark jet  sample contains 11\% $b$'s  and 33\% $c$'s. All
other  results are corrected  from their original  values  to account for the
same quark mixture. In  figure~\ref{fig:proc_mult}.a the quark and gluon  jet
multiplicities are shown as a function of the jet energy and compared to {\sc
Jetset}.  Available  results  from  LEP\cite{opal_qg,aleph_qg,delphi_qg}  and
PEP\cite{hrs_qg} using  symmetric   and   non-symmetric  configurations   are
entered.   In   addition  to these data    figure  \ref{fig:proc_mult}.b also
includes measurements  from SLD\cite{sld_qg}. The value given   for $r$ at 10
GeV and 15 GeV is calculated using the quark jet  multiplicities from HRS and
the gluon jet multiplicities from DELPHI.

Evidence for an energy  dependence of $r$ is   again observed when using  all
these data. The measured increase is
\[
\frac{\Delta r}{\Delta E} = 
(+97 \pm 21\ \mbox{(stat.+syst.)})\cdot10^{-4}\  \mbox{GeV}^{-1},
\]
representing a $\sim$5$\sigma$ effect.

The measured value of $r$ remains  systematically lower than the {\sc Jetset}
prediction over the whole energy range, having an average value of
\[
r= 1.23\pm0.01\   (stat.)  \pm0.03\  (syst.),
\]
corresponding to an average energy of $\sim$22 GeV. This ratio can be further
expressed as
\[
r_{uds}= 1.30 \pm0.01\ (stat.) \pm0.04\ (syst.), 
\]
if $r$ is computed only for the light quarks $u$, $d$ and $s$, extracting the
$b$ and $c$ quark contribution  to  the quark  jet multiplicity according  to
\cite{delphi_bm,opal_bm,sld_bm}.

\section{Inter-jet coherence, the string effect}
The measurements of the energy and particle  flow in the inter-jet regions of
three-jet $q\bar{q}g$ and $q\bar{q}\gamma$ events represent an important test
on  the non-abelian  gauge nature of   QCD. Inter-jet  coherence  effects are
expected to be seen in data according to the  Local Parton Duality hypothesis
(LPHD). Again  the possibility to  perform either qualitative or quantitative
tests depend on the available theoretical predictions\cite{string_calcul} and
on the selected experimental data sample.

Using Mercedes  events  DELPHI\cite{delphi_qg} measures an asymmetry   of the
particle      flow    between  the $qg$        and     $q\bar{q}$ regions  of
$R_g=N_{qg}/N_{q\bar{q}}=2.23\pm0.37\ \mbox{(stat.+syt.)}$, proving  that the
string effect is present in fully symmetric three-jet event configurations in
which all jets  have the same energy.  This  rules out those arguments  which
explain the string effect on the basis of only kinematic considerations.

\begin{figure}[hbt]
\begin{center}
\mbox{\epsfig{file=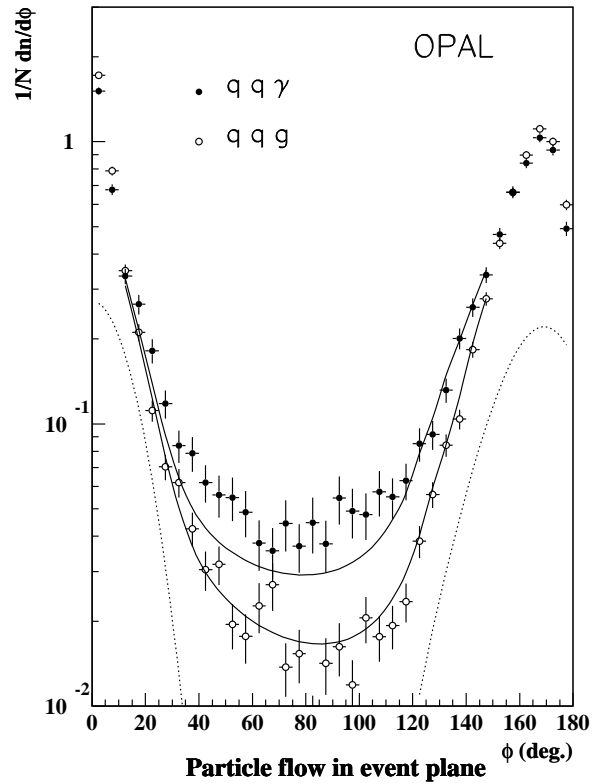,width=8.0cm}}
\end{center}
\caption{Charged particle flow in the inter-jet $q\bar{q}$
region of $q\bar{q}g$ and 
$q\bar{q}\gamma$ events as compared to the MLLA 
prediction. The dotted line  represents  the  estimated intra-jet 
contribution.}
\label{fig:opal_figura}
\end{figure}

A cleaner verification  of the string  effect is then  evident when comparing
the particle flow in  the  $q\overline{q}$ region of $q\overline{q}g$  events
with   the  corresponding  region  in   $q\overline{q}\gamma$ events.  Figure
\ref{fig:opal_figura} shows the  OPAL results\cite{opal_string} for three-jet
events using all topological configurations.  The angular separation  between
jets was  normalized   to allow   a proper   comparison  and   the   obtained
distribution was fitted considering  a perturbative QCD contribution as given
by the MLLA approach and a non-perturbative intra-jet contribution assumed to
be gaussian. The results of the fit are in reasonable agreement with data.

The DELPHI approach\cite{delphi_qg} is different to that from OPAL as it uses
symmetric $q\bar{q}g$  and $q\bar{q}\gamma$ three-jet  events.  The  selected
configuration corresponds to Y events  which have a $q\bar{q}$ jet separation
of   $150^\circ\pm10^\circ$. The  measured     ratio  of the particle    flow
corresponding    to   this     region   in   both    type     of    events is
$R_q=N_{q\bar{q}}(q\bar{q}g)/N_{q\bar{q}}(q\bar{q}\gamma)=0.58\pm0.06\
\mbox{(stat.+syt.)}$, compatible with the   MLLA prediction of 0.60. This  is
the first time  in which inter-jet   coherence effects are measured to  occur
according to the  perturbative QCD prescription. This is   mainly due to  two
important  aspects  of the  analysis, the  highly  pure sample of  gluon jets
(96\%) which enhance  the effect and the  large angle separation between jets
which minimizes non-perturbative contributions.

\begin{table}[htb]  
\caption{Particle flow ratios measured by the LEP experiments 
using $q\bar{q}\gamma$ events and the characteristics of each analysis.} 
\begin{center}
\begin{tabular}{|l|c|c|c|}
\hline
\hline
\multicolumn{1}{|c|} {Experiment}                           & 
\multicolumn{1}{c|} {Topology}                        & 
\multicolumn{1}{c|}{$R_q$} &
\multicolumn{1}{ c| } {Gluon id.}                 \\
\hline \hline
ALEPH  &  Y   & $0.75\pm0.07$ & Lepton \\
DELPHI &  Y   & $0.58\pm0.06$ & Vertex \\
DELPHI &  Y   & $0.68\pm0.07$ & Lepton \\
L3     &  All & $0.79\pm0.06$ & Lepton \\
OPAL   &  All & $0.71\pm0.03$ & Energy \\
\hline
\hline
\end{tabular}
\end{center}
\end{table}

In    table       1      all    LEP   measurements      of      the    string
effect\cite{delphi_qg,opal_string,l3_string,aleph_string}                with
$q\bar{q}\gamma$ events  are summarized. The  differences among these results
can be understood in terms of the applied gluon  selections and the three-jet
configurations used in the analyses.

\section{Fragmentation    functions}  
\begin{figure}[htb]
\begin{center}
\mbox{\epsfig{file=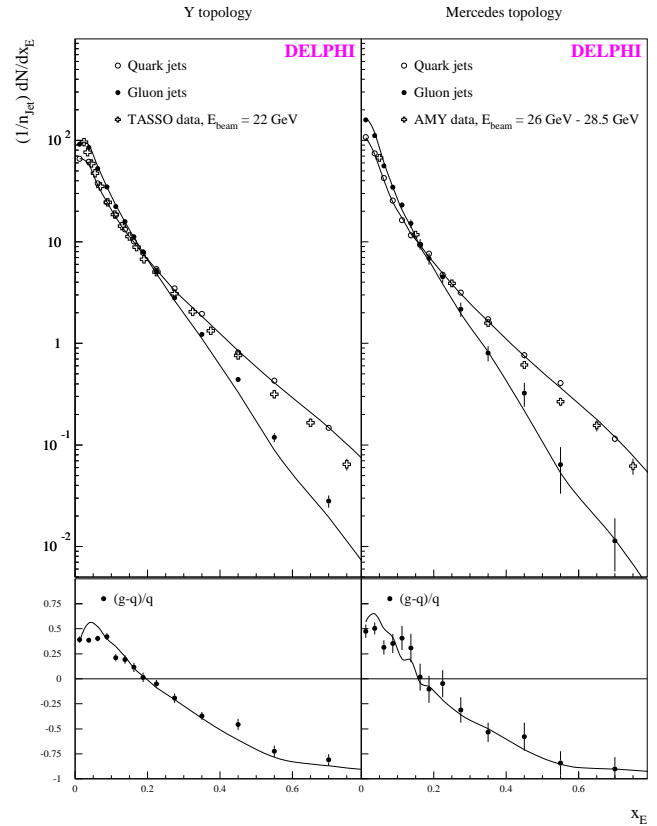,width=9.0cm}}
\end{center}
\caption{Fragmentation functions for quark   and gluon jets selected   from
symmetric three-jet  events. The solid  lines are the  corresponding 
{\sc Jetset}
predictions as tuned using DELPHI data.}
\label{fig:proc_scaling}
\end{figure}

The distribution of  the energy fraction  carried by the generated particles,
$x_E =E_{part}/E_{jet}$, is known as the  fragmentation function. LEP studies
using  symmetric Y  three-jet  event   configurations\cite{opal_qg,delphi_qg}
demonstrate that the gluon jets contain a softer particle spectrum than their
quark   jet counterparts of   equivalent  energy.   DELPHI  also includes  an
analyses   on  fully symmetric three-jet    events  (Mercedes).  A comparison
between the obtained distributions at both energy points, 24 GeV for Y and 30
GeV for  Mercedes events, shows that  the increase  in the multiplicity takes
place at small momentum  whereas  the inclusive  spectra decreases for  large
$x_E$, indicating  thus  the  presence  of  scaling  violations (see   figure
\ref{fig:proc_scaling}). The measured suppression is   larger for gluon  jets
than for quark jets by a factor $2.4\pm0.5\ \mbox{(stat.)}$, which is in nice
agreement with the lowest order QCD prediction of $\sim$2.5.

\section*{Acknowledgments}

We thank  all colleagues from the  LEP experiments that helped  us collecting
information for  the elaboration of these proceedings.  We especially like to
mention K. Hamacher, O. Klapp and P. Langefeld for many useful discussions.

\end{document}